\documentstyle[spie]{article}
\input{psfig}
\newcommand{\mic}{\,{\rm \mu m}}
\title{In.XS: project for a future spaceborne hard X-ray all-sky survey}
\author{Philippe Marty\supit{a},
        Juho Schultz \supit{b},
        Clemens Bayer \supit{c,d},
        Alexander Fritz \supit{c,e},\\
        Martin Netopil \supit{c},
        Walter Nowotny \supit{c},
        Michael Carr \supit{f},
        Carlo Ferrigno \supit{g},\\
        Christophe Jean \supit{h},
        Walter Koprolin \supit{c},
        Jesper Rasmussen \supit{i},
        Laura Tanvuia \supit{c},\\
        Ivan Valtchanov \supit{j},
	Marcos Bavdaz \supit{k},
        Rudolf Much \supit{k},
        Arvind Parmar \supit{k}
\skiplinehalf
\supit{a} Institut d'Astrophysique Spatiale
\\Universit\'e Paris-Sud - B\^at.121, F-91405 Orsay cedex
\\
\supit{b} Helsinki Observatory
\\University of Helsinki, P.O. Box 14, FIN-00014 T\"ahtitorninm\"aki
\\
\supit{c} Institut f\"ur Astronomie
\\Universit\"at Wien, T\"urkenschanzstr. 17, A-1180 Wien
\\
\supit{d} Inst. of Astronomy, University of Cambridge, Madingley Road, CB3 0HA Cambridge, UK
\\
\supit{e} G\"ottingen Obs., Universit\"atssternwarte, Geismarlandstr. 11, D-37083 G\"ottingen
\\
\supit{f} Dunsink Observatory
\\Shool of Cosmic Physics, Castleknock Dublin 15, Ireland
\\
\supit{g} S.R.O.N.
\\Sorbonnelann 2, NL-3584 CA Utrecht
\\
\supit{h} Institut d'Astrophysique et de G\'eophysique
\\Universit\'e de Li\`ege, 5 av. de Cointe, B-4000 Li\`ege
\\
\supit{i} Astronomical Observatory
\\University of Copenhagen, Juliane Maries Vej 30, DK-2100 Copenhagen
\\
\supit{j} C.E.A. / Service d'Astrophysique
\\Orme des Merisiers, Bat. 709, F-91191 Gif/Yvette cedex
\\
\supit{k} E.S.T.E.C. / Astrophysics division
\\PO.BOX 299, Keplerlaan 2, NL-2200 AG Noordwijk
}
\authorinfo{Further author information: (Send correspondence to P.M.)
\\ P.M.: E-mail: philippe.marty@ias.u-psud.fr}
\begin{document}
\maketitle
\begin{abstract}
The latest all-sky survey in hard X-ray band was performed by the
HEAO-1 satellite ($13-80 keV$) with an angular resolution of $24 \times 48
arcmin$\cite{Lev84}. A diffuse hard X-Ray background
(HXB) was detected between $3$ and $50 keV$\cite{Mar80}. The
main scientific goal of In.XS is to resolve a large fraction of this
HXB into individual sources.

As no distortion by Compton up-scattering is seen in the spectrum of
the microwave background\cite{Mat94}, the hard X-ray
background is believed to be mainly due to point sources.
``Type I'' Active Galactic Nuclei (AGN) have softer X-ray spectra
than the hard X-ray background, so other sources must be considered, like
faint ``Type II'' or ``absorbed'' AGN. These could be distinguished through hard
X-ray spectroscopic or hardness ratio observations.

Here we present In.XS - a mission concept designed to conduct
the first imaging all-sky hard X-ray ($2-80 keV$)survey. The angular resolution
of nearly $1 arcmin$ and good sensitivity at high energies is 
provided by the latest multilayer focussing mirrors, with 
semiconductor-based (GaAs) arrays of detectors. We
also describe the mission operations, and how the all-sky survey
will be complemented by follow-up pointed observations of selected
fields.
The good angular resolution will allow correlations and
identification with objects seen at other wavelengths. In addition, 
since a large fraction of the Type II AGN luminosity is emitted in the 
hard X-ray band, this survey will provide a large unbiased sample
of the AGN population. This may provide constraints on AGN evolution
through the possible observation of a turnover in deep
field source statistics.
\end{abstract}

\section{INTRODUCTION}

Since the first (1964) observational proof of the existence of a cosmological radiation backround (CRB)
in radiowaves\cite{PW64}, the question has been raised of its extension along the
electromagnetic spectrum and its origin.

If the radiowave background, as well as its extension into microwaves later (1990) observed by the COBE satellite,
is now widely admitted to originate around the time of recombination (z=1000), other diffuse emissions have
been detected at different wavelengths and their origins still remain hypothetical.
   A strong background has indeed been discovered in the far-infrared data from COBE\cite{Pug96}.
The energy contained between wavelengths $6 \mic$ and $1 mm$ is about twice the energy contained in the ultraviolet-visible
part of the spectrum\cite{Gis99} and is by far the most dominant component of the CRB after the microwave background
itself.

\subsection{The X-ray Background (XRB)}

In 1967, isotropic X-ray emission was evidenced after rocket experiments in the $4 - 40 keV$ band with
a somewhat powerlaw-like spectrum\cite{Sew67} and was hence primarily thought to be due to inverse Compton scattering of microwave
photons on electrons in a hot intergalactic medium\cite{Bre67}. But as no strong distortion by Compton up-scattering is seen in
the spectrum of the microwave background\cite{Mat94}, this scenario shall be disregarded.

While $75\%$ of the soft XRB ($0.5 - 2 keV$) has been since resolved into individual sources\cite{Has98} by ROSAT,
as well as $75\%$ of the mid XRB ($2 - 10 keV$) by the Chandra satellite\cite{Mus00}, this is not true for the hard
X-ray component ($10 - 100 keV$) of the background.

Apart from invoking exotic models
there are two main sources of radiation in the universe in the period
following recombination during which structures form: gravitational energy of contracting or accreting objects and
nucleosynthesis in stars of H and He into heavier elements. In the first case, only Black Holes
can radiate a substantial fraction of their rest mass during formation or accretion processes\cite{SR98}. In the
second case, starburst in galaxies can radiate five times as much energy as galactic black holes\cite{Lag99}.

\subsection{Active Galactic Nuclei (AGN)}

\begin{figure} [h]
\begin{center}
\hspace{0cm}
\psfig{figure=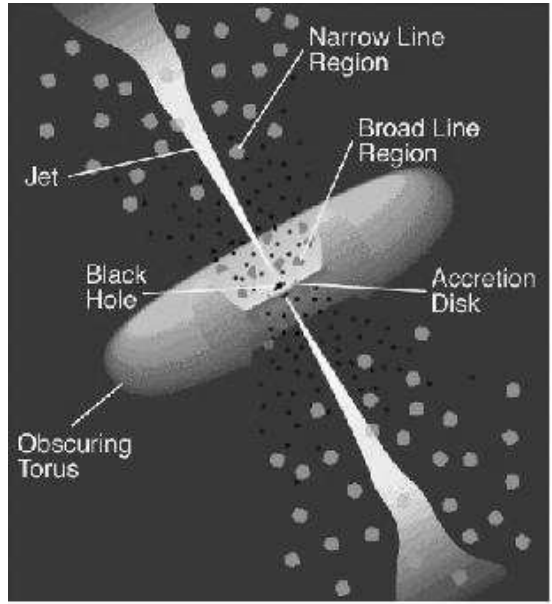,height=8cm}
\caption{\label{agn} Sketch of the Unified AGN model. Type 1
are those seen along the direction of the jet while Type 2 are
those seen from the torus side.}
\end{center}
\end{figure}

Individual sources powering softer components of the XRB were demonstrated to be mainly Seyfert galaxies\cite{Sch98}, not surprisingly
of that kind presenting a strong evidence for an accreting supermassive black hole at their center.
In the framework of the AGN unified scheme (Fig.\,\ref{agn}), these galaxies are refered to as ``Type 1 AGN''.
The problem is that their spectra
do not account for the bump in the harder XRB components, so that other sources must be considered\cite{Mad94}.

Nevertheless, a large ($> 90\%$) fraction of the hard X-ray background (XRB)
spectral intensity could be well reproduced by the combined emission of unobscured (type 1) and obscured (type 2) Seyfert galaxies
and quasars with a distribution of absorption column densities and luminosities\cite{Com95}. The key ingredient of this model 
is the presence of obscured AGN whose emission in the traditional $2 - 10 keV$ band is blocked by
the dust torus but allows to fit the bump around $30 keV$.

It is likely that the energy density of the absorbed AGN which would peak in the soft X-rays is reradiated in the far-infrared
and assuming that the $30 keV$ peak of the XRB provides an unobscured estimate of the integrated AGN energy density, the
obscured AGN responsible for the hard XRB could account for no less than $15\%$ of the far-infrared background\cite{Fab98}
and as much as $50\%$\cite{Man99}. On the other hand, it is shown on the basis of spectroscopy (carried out with ISO) of luminous and ultraluminous IR galaxies, that the fraction of the energy coming from the central AGN is in general smaller than the fraction coming from starburst activity but increasing with luminosity\cite{Gen98}.

The far-infrared and the hard X-ray bands should hence be the best two regions of the spectrum to select AGN in a manner insensitive
to their level of obscuration.

\section{In.XS MISSION CONCEPT}

The latest all-sky survey in hard X-ray band was done by the HEAO-1
satellite. Its A-2 instrument detected 72 point sources\cite{Lev84} to the limit
of $13.3 mCrab$ in the $13 - 80 keV$ band with an angular resolution of
$24 \times 48 arcmin$. A diffuse hard X-ray background\cite{Mar80} of 
$5.13 \cdot 10^{-9} erg \cdot keV^{-1} \cdot sr^{-1} \cdot cm^{-2} \cdot s^{-1}$
at $10 keV$ was also detected.

In addition, there is currently an ongoing effort to re-analyse BATSE data with a 
novel analysis technique and perform a sensitive all sky survey
in the $20 - 300 keV$ range.
The 3 sigma flux sensitivity for a survey performed using 50 days of continuous data 
is $\sim 15 mCrab$ with an angular resolution of $1 deg$\cite{Sha00}.

The main scientific goal of the International X-ray Surveyor (In.XS) is to resolve a large fraction of this diffuse hard X-ray background
into individual sources, mainly AGN hidden in other wavelength bands. A comparison between the high performance of In.XS 
with that of other missions is made in Tab.\,\ref{missions}.

\begin{table} [h]
\caption[]{\label{missions} Comparison to other missions.}
\begin{center}
\begin{tabular}{|l|c|c|c|c|}
\hline
{\bf Mission/Instrument} & {\bf FOV}       & {\bf Resolution} & {\bf Energy range} & {\bf Sensitivity}                   \\
(* not yet launched)     & (degrees)       & (arcmin)         & (keV)              & (mCrab) (*not all-sky survey)       \\
\hline
In.XS *                  & $0.5$           & $1$              & $2 - 80$           & $5 \cdot 10^{-2} - 5 \cdot 10^{-4}$ \\
\hline
HEAO-1 A-2               & $1.5 \times 20$ & $24 \times 48$   & $13 - 80$          & $13.3$                              \\
Granat ART-P             & $1.8$           & $5$              & $4 - 60$           & $1$ * (in 8 hours)                  \\
BATSE                    &                 & $60$             & $20 - 300$         & $15$                                \\
ROSAT                    &                 & $\simeq 0.5$     & $0.5 - 2$          & $10^{-4} - 10^{-5}$                 \\
XMM-Newton or Chandra    & $0.5$           & $\simeq 0.1$     & $0.5 - 10$         & $10^{-3} - 10^{-4}$                 \\
INTEGRAL JEM-X * (2002)  & $4.8$           & $3$              & $3 - 35$           & $10$ * (in 15 min)                  \\
SWIFT * (2003)           & $80$            & $22$             & $10 - 150$         & $1.3$                               \\
ISS EXIST * (2007)       & $160 \times 40$ & $5$              & $10 - 100$         & $0.05$                              \\         
SRG MARTLIME *           & $6$             & $9$              & $5 - 150$          & $1$ * (in 24 hours)                 \\
CHIP *                   & $60$            & $26$             & $2-100$            & $1$ (in 1 hour)                     \\
\hline
\end{tabular}
\end{center}
\end{table}

\subsection{Scientific Goals}

With a detection limit of
$10^{-12} erg \cdot s^{-1} \cdot cm^{-2}$
in the $10 - 70 keV$ band, approximately 20000 new sources are expected,
most of them AGN. This source count estimate is derived from a
$LogN-LogS$ relation with $-3/2$ slope, consistent with HEAO-1\cite{Pic82} and
ASCA\cite{Ued99} observations. A fraction of $1-10\%$ of the new
sources would be galactic, most of them accreting X-ray pulsars.
Crude variability estimates would be obtained, as the satellite will scan the sky twice,
with six months between 2 exposures of $60 sec$ each (cf. Sec.\,\ref{orbit}.).

The planned angular resolution of nearly $1 arcmin$ allows for correlation
with other deep X-ray surveys\cite{Has98} or infrared background surveys\cite{Pug96}, for
easier identification of counterparts in other wavelengths and provides a limited
level of sources confusion.

The observed hardness ratios of sources would allow crude
division between AGN type 1 and type 2. The ratio of AGN I to AGN II would hence
allow estimates of the average dust torus size and
AGN I/II ratio versus redshift would give some constraints on
dust torus evolution. As a large fraction of the AGN luminosity
is emitted in the hard X-ray band, and only a few small classes of other
sources are seen in this band, a good population synthesis of AGN
would be easiest to obtain from a large sample of AGN observed
in hard X-rays.

\subsection{Secondary Goals}

The main result of the first mission phase, the all-sky survey,
will be a large source catalogue.
In the second phase of the mission pointed observations will be conducted, 
whereby interesting targets could be selected from the source catalogue
compiled during the first phase. This can also include raster scans of 
selected regions and deep field observations
of a few very interesting regions (e.g. Galactic Centre and Lockman
Hole). The deeper observations are needed to
calibrate the hardness ratio - AGN type relation by obtaining a
sufficient sample of AGN with known types and better spectra, and should
lead to about $300 sources \cdot deg^{-2}$ in the $10 - 40 keV$ range
with a detection limit of $3 \cdot 10^{-14} ergs \cdot cm^{-2} \cdot s^{-1}$ in $100 ksec$.

The AGN evolution models would be constrained if 
the turnover of the $LogN-LogS$ relation is seen in the deep field observations.
Since the sources of the all-sky catalog contribute to roughly half of 
the diffuse X-ray background, there is a good chance to reach the turnover
point in the deep field observations. 

Furthermore, in addition to the test of the unified model of AGN provided by the correlation between
X-rays and existing IR surveys (ISO FIRBACK and ELAIS), the combination of infrared probes of dust conditions
(to be conducted in the forthcoming years with SIRTF and FIRST-Herschell), together with radio and hard X-ray
data probing supernovae and X-ray binaries activity, should lead to major
constraints on starburst evolution models in galaxies. Also, having both
IR and X-ray photometry for these objects should enable a detailed study of how X-ray photons are reprocessed
by the dust torus, and facilitate the estimation of the intrinsic spectrum of these X-rays.

Finally, this all-sky survey should be compared with its submillimeter counterpart to be realized by
the Planck satellite by the end of the decade and which should provide us with maps of the Sunyaev-Zel'dovich (SZ)
effect, Compton recoil of the CRB photons on the hot electronic gas of clusters of galaxies (ClG), and deep ClG
catalogs. This comparison could put further constraints on the possible diffuse component of the XRB and
possibly maps of the Hubble parameter $H_0$, since the SZ effect is not dependant on the redshift of the source object.




\section{In.XS INSTRUMENT PAYLOAD}

The scientific goals, as outlined in previous section, define the scientific mission
requirements to be achieved as listed in Tab.\,\ref{prereq}.

\begin{table} [h]
\caption[]{\label{prereq} Summary of performance to be achieved.}
\begin{center}
\begin{tabular}{|c|c|c|c|c|}
\hline
{\bf FOV}       & {\bf Resolution} & {\bf Energy range} & {\bf Sensitivity} & {\bf Energy Resolution} \\
(deg)           & (arcmin)         & (keV)              & (mCrab)           & (keV)                   \\
\hline
$0.5$           & $1$              & $2 - 80$           & $5 \cdot 10^{-2}$ & $\le 1$ at $40 keV$     \\
\hline
(to allow scanning &(to allow &(to explore &(to allow unbiased &(to discriminate \\
 all-sky in 6 & correlations and & absorbed regions & sampling of AGN & between AGN \\
 months from LEO) & cross-identifications) & of X-ray spectrum) & population) & types spectra) \\
\hline
\end{tabular}
\end{center}
\end{table}

\subsection{X-ray Telescopes}

Since 1978, grazing incidence mirrors using Bragg theory have been the 
working horse of X-ray astronomy, but these mirrors were limited in
energy range. In 1993, this limitation was alleviated from about $4$ up to $10 keV$, but at the
cost of spatial resolution for a (from $4 arcsec$ for HEAO-2 down to $3 arcmin$
for ASCA), since the Bragg condition roughly implies
${{\theta} \over {\delta \theta}} = {{E} \over {\delta E}}$, where
$\theta$ is the critical incidence angle, $\delta \theta$ half the FOV, $E$ the considered energy
and $\delta E$ the energy resolution. These grazing mirrors, coated with atomic
monolayers (to improve reflectivity) and designed as a
combination of paraboloid and hyperboloid surfaces
(instead of simple conical surfaces) to improve spatial resolution (XMM-Newton now reaches
about $5 arcsec$ spatial resolution) and reduce chromatic aberrations,
are still in use in modern X-ray space observatories.

But it has been demonstrated that a multilayer coating enhances the reflectivity
beyond the critical angle\cite{Yam97}, up to $100 keV$, since the photoelectric
absorbtion is less effective in the multilayer structure, so that the Bragg
condition still is satisfied over the broad band. Alternance of Pt/C layers is
hence proved to be the best combination, in addition to Pt monolayer overcoating.

\begin{figure} [h]
\begin{center}
\hspace{0cm}
\psfig{figure=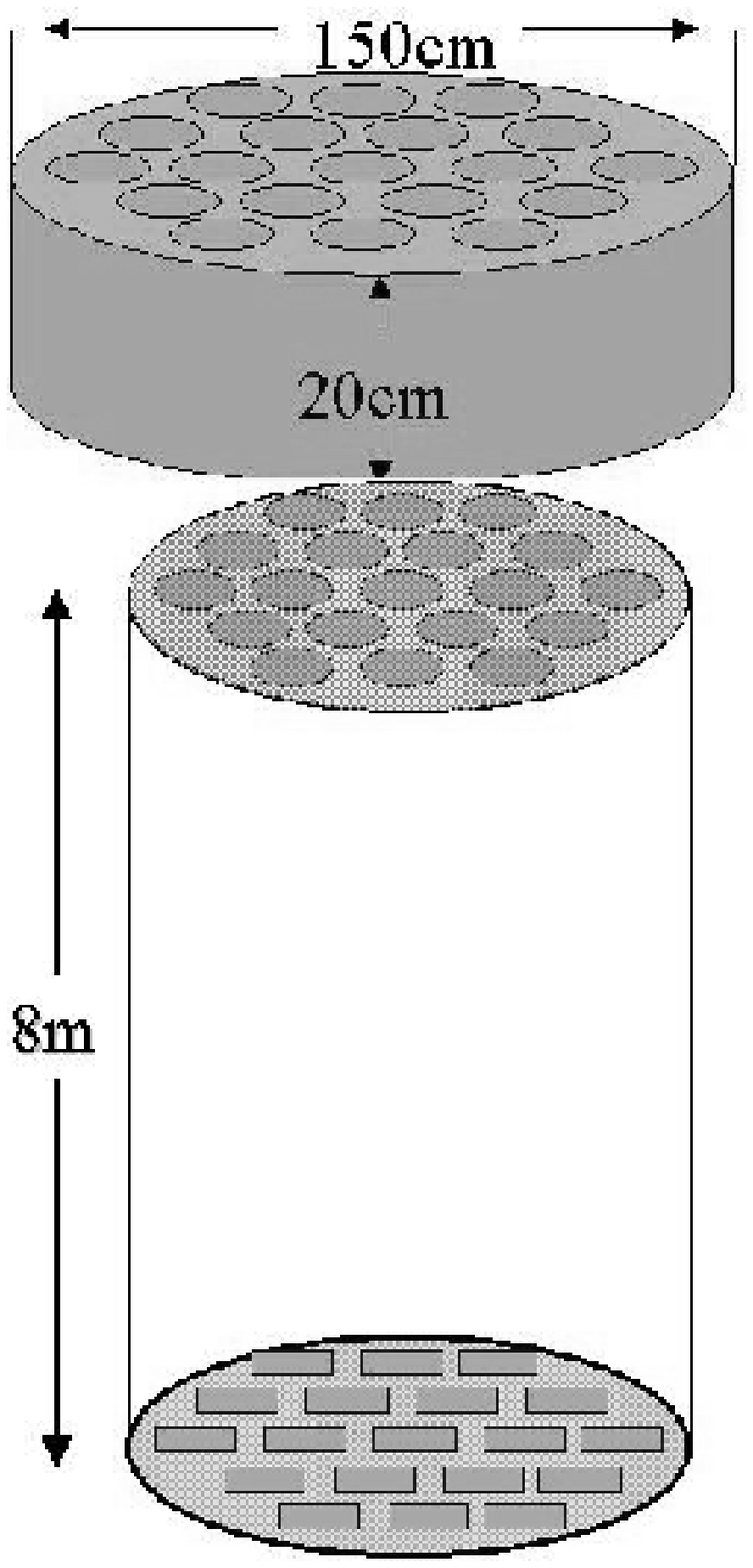,height=8cm}
\psfig{figure=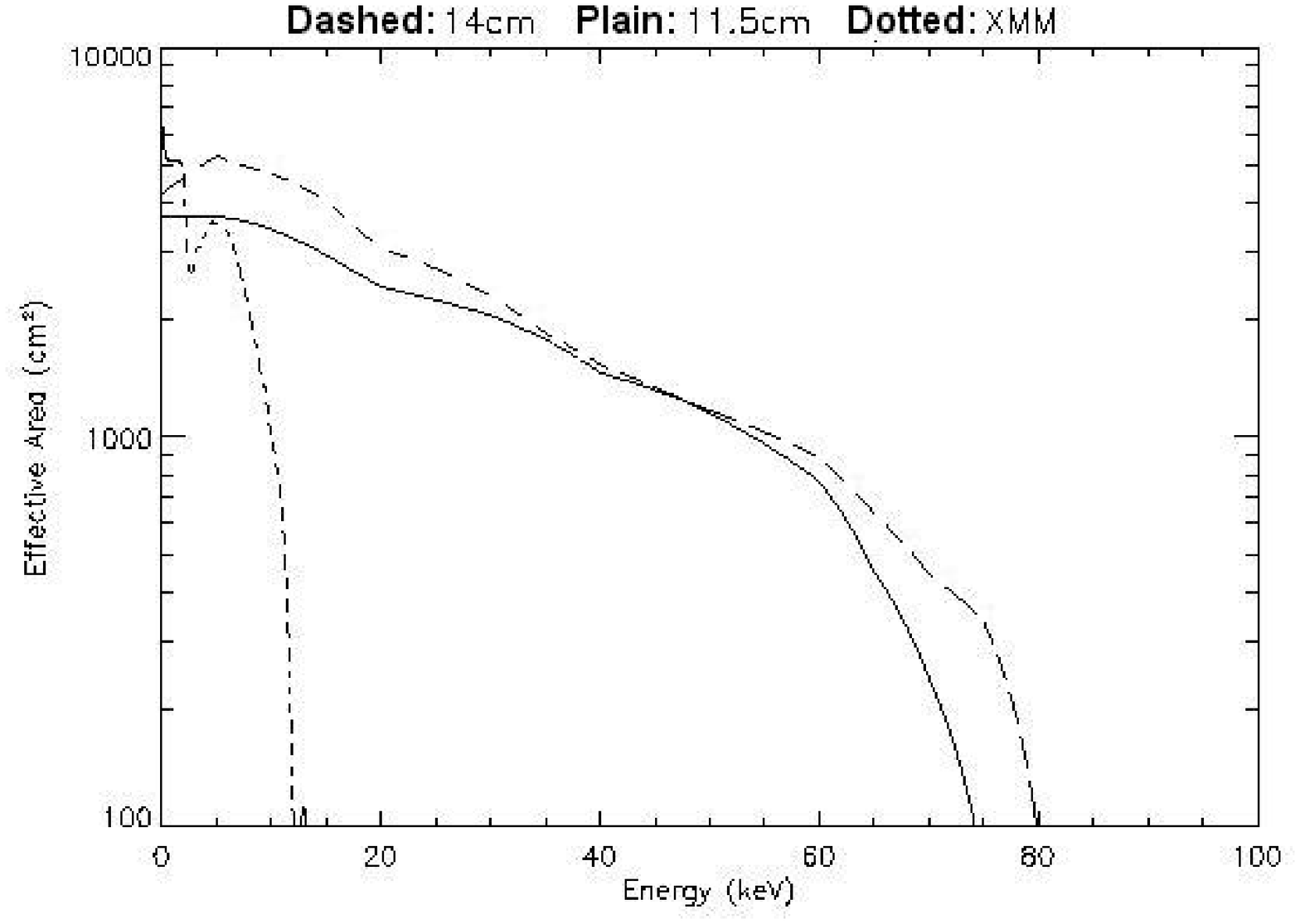,height=8cm}
\caption{\label{mirror} (LEFT) Sketch of the In.XS mirror platform (left top) and of the entire
telescope assembly (left bottom).}
{(RIGHT) Plot of the In.XS 19 telescopes total effective
area for 2 different radii of each modules, compared to that of the XMM-Newton 3 telescopes.}
\end{center}
\end{figure}

In order to meet the sensitivity requirements, an effective area of $1500 cm^2$
at $40 keV$ is needed. This will be achieved for a $8 m$ focal length by $19$ mirror
units, each made of about 100 multi-nested multilayers ($40$ Pt/C layer pairs)
shells (Fig.\,\ref{mirror}). Each unit has a diameter of $30 cm$ and a length of $20 cm$.

Finally, applying the former Bragg relation to our science requirements, we find that
it should be difficult to achieve more than $0.2 deg$ FOV without reducing
performance. However, it is reported that with a higher (9th) polynomial degree in the 
``Wolter I'' approximation of the mirror shape, wider FOV could be reached\cite{Cit99}.

This assembly constitutes a major improvement as compared to what is planned for InFOC$\mu$S
and Constellation-X designs.

\subsection{X-ray Solid State Detectors}

In a survey of the available detector technologies (cf. Tab.\,\ref{detec}),
GaAs compound semiconductors were found to fit our science requirements best.

\begin{table} [h]
\caption[]{\label{detec} Comparison with various detector types.}
\begin{center}
\begin{tabular}{|l|c|c|c|c|}
\hline
{\bf Type}      & {\bf Missions}  & {\bf Energy range} (keV) & {\bf Comments}                    \\
\hline
GaAs array      & In.XS           & $0.1 - 200$              & pixel size $< 1 mm$,              \\
                &                 &                          & $\delta E < 1 keV$,               \\
                &                 &                          & extremely uniform response        \\
\hline
CdZnTe array    & InFOC$\mu$S, CHIP,  & $1 - 100$            & pixel size $> 1 mm$,              \\
                & Constellation-X &                          & $\delta E > 1 keV$,               \\
                &                 &                          & discrepancies from pixel to pixel \\
TlBr array      &                 & $10 - 200$               & $\delta E > 1 keV$                \\
Si(Li) diode    &                 & $< 40 keV$               & $T < -50^{\circ}C$                 \\
Si CCD          & ASCA, XMM, AXAF & $< 20 keV$               &                                   \\
Gas Prop. Coun. & MARTLIME, ROSAT & depends on gas           & big pixels \ldots                 \\
$\mu$Calorimeter& rockets         & $< 10 keV$               &                                   \\
Bolometer       &                 &                          & must be cooled \ldots             \\
Diamond         &                 & $< 1 keV$                &                                   \\
\hline
\end{tabular}
\end{center}
\end{table}

GaAs arrays work perfectly in our required $2 - 80 keV$
energy range. They provide a sufficient spectral resolution ($300$ to $700 eV$)
already at room temperature, which prevents usage of
costly onboard cryogenics (at low earth orbits, spacecraft has to endure
substantial thermal variations). An improved energy resolution (by a factor of 2)
can be achieved with simple passive cooling down to $-35^{\circ}C$\cite{Bav99}.

The best absorption layer thickness should be around $400 \mic$ for pixels
size comprised between $0.2$ and $1 mm$ and a
resulting quantum efficiency of $70$ to $5\%$ in our energy range\cite{Owe00}.

It is reported (Metorex Inc., http://www.metorex.fi, June 2000) that
industrial GaAS arrays of $32 \times 32$ pixels are now available, which
is enough to sample our $30 arcmin$ FOV at spatial resolution
consistent with mirror PSF ($1 arcmin$). Given the telescope characteristics,
a detector array of $8 \times 8 cm$ is needed to fit the focal
plane of one unit. However, GaAs wafers yet come no larger than
$5 \times 5 cm$. Therefore, $4$ square arrays of $4 \times 4 cm$ each should 
be considered to provide a total of $64 \times 64$ pixels behind each telescope unit.
The needed pixel size would then be around $1 mm$, corresponding to
$\sim 28 arcsec$ on the sky (Shannon sampling of the PSF).

\subsection{Readout Electronics}

GaAs detectors can be equipped by individual wire bonding
(4096 wires per detector unit in our case), thus allowing
spectro-imaging by independent energy determination in each pixel of the image.
Each pixel is then preamplified (FET) and a global spectro-amplifier feeds
a $12 bits$ digitizer (thus allowing us to divide our $80 keV$ energy
range into 4096 channels of $20 eV$)\cite{Owe00}. These are classical
space qualified technologies and the electronic noise is no more
than $10^{-4} cnts \cdot cm^{-2} \cdot s^{-1}$, that is
just less than our expected cosmic background at $100 keV$.

The readout time cycle of the electronics is
of the order of a few microseconds, which prevents smearing of the PSF.
The sky scanning rate of the satellite is $4 arcmin \cdot s^{-1}$
(cf. Sec.\,\ref{orbit}), equivalent to $9$ detector pixels per second.
Smearing of the PSF over several pixels is hence avoided with
a readout cycle faster than $1/9 sec$, which is easily fulfilled.

The encoding of data must be considered through an event list (science data)
and housekeeping. Each event should be characterized by its position (telescope
number: $5 bits$, pixel coordinates $2 \times 6 bits$), energy ($12 bits$)
and timestamp (a typical long integer on $24 bits$), for a total of $53 bits$.
The signal may vary from $0.05 cnts/s$ ($30 keV$ X-ray background flux on our telescopes)
up to probably $500 cnts/s$ for the brightest objects (one hundred times less than
the pile-up limit thanks to individual wiring electronics), for a data rate
of $2.5 bits/s$ to $25 kbit/s$. It seems reasonable to assume an average
value of $3.2 kbits/s$ dominated by the standard housekeeping data rate.

\subsection{Instrument Calibrations}

A first run of ground based calibrations could be undertaken on monochromatic
X-ray sources with large housing capability for effective area and
vignetting measurements of mirror modules (like the Panter station
in Germany); the detectors could be hosted by white synchrotron facilities
(like Bessy in Germany or the new SoLEIL in France) for quantum efficiency,
flat field and spectral resolution measurements. One should particularly take
care of variation of these
parameters with temperature, since
the superstructure may have to endure significant thermal fluctuations along its
low earth orbit (cf. Sec.\,\ref{orbit}) and that may produce strong variations
in the mirror performance and between each detector preamplifying chain.

Onboard radioactive sources ($^{55}Fe$ and $^{241}Am$) as well as well known astrophysical
sources (3C273 Quasar, Crab Nebula, etc.) would provide inflight calibrations.
Common targets during the second phase (pointed observations) should also be considered for 
cross-correlations with other observatories like XMM-Newton or INTEGRAL. A filter wheel
in front of each detector with different positions (``open'', ``filter'', ``closed'')
would allow adjusting incident flux towards limited damage flux level. 
In addition, this would ensure protection of the detectors against radiation damage during strong
solar flares.

\section{In.XS SPACECRAFT AND OPERATION}

\subsection{\label{orbit} Spacecraft Design and Properties}

The optimal choice of orbit for our scientific requirements would be a
circular low earth orbit (LEO) at an altitude of about $600 km$. This choice of
orbit allows us to keep the background at the lowest possible level,
while at the same time avoiding the radiation belts. Furthermore this choice
keeps the operational costs of the mission (launch \ldots) within reasonable limits.
The inclination of the orbit will be 7 degrees. An equatorial orbit is possible, 
but is not necessary for our purposes.

\begin{figure} [h]
\begin{center}
\hspace{0cm}
\psfig{figure=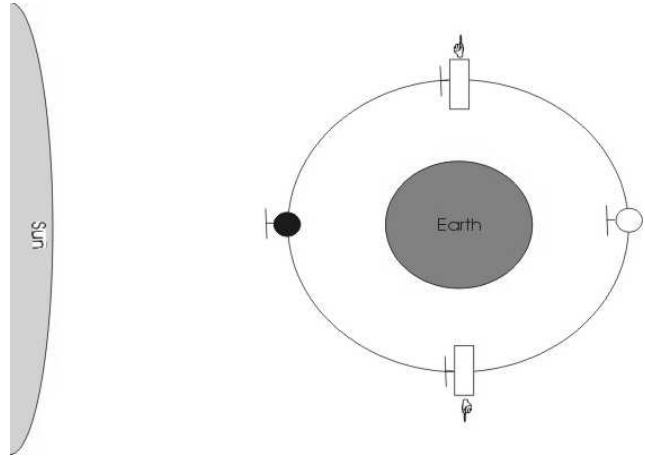,height=6cm}
\caption{\label{period} Sketch of the In.XS orbital phases.}
\end{center}
\end{figure}

By choosing a rotational period of the spacecraft around its transversal axis such that it
equals its period of revolution $P_{rot} = P_{rev} \simeq 97 min$ (Fig.\,\ref{period}),
we enable the spacecraft to scan the entire sky within 6 months, while
at the same time pointing away from the earth at all times. The same result could have
been achieved with a polar orbit while constantly pointing away from the earth and the
sun, but this would have meaned suffering from polar aurorae radiation.

The proposed minimum duration of the first part of the mission should
therefore be no less than 1 year (ideally 1.5 years including Com/Cal/PV Phases) to allow for redundancy, instrument
failures and in order to cover those FOV obstructed by the Moon in the first run.

It should be noted that, although our choice of orbital motion does
allow the solar panels to be pointing towards the sun at all times, the
low altitude of the orbit leads to a significant amount of sky
shadowing of approximately $130^\circ$, resulting in an eclipsing time of 36
minutes per orbit. Therefore we have to account for batteries and data
storage on board the spacecraft.

We estimate the total mass of the spacecraft to be approximately $1.8 tons$,
including $700 kg$ for superstructure, $500 kg$ for science payload
and $200 kg$ of monopropellant necessary for altitude
corrections and de-orbiting at the end of the mission. Due to the
relatively short overall lifetime of the mission (3 to 6 years), further
orbit maintenance (i.e. altitude correction of an annual loss of about $30 km$) is not vital; this
allows us to economize propellant and keep the total mass down.

\begin{figure} [h]
\begin{center}
\hspace{0cm}
\psfig{figure=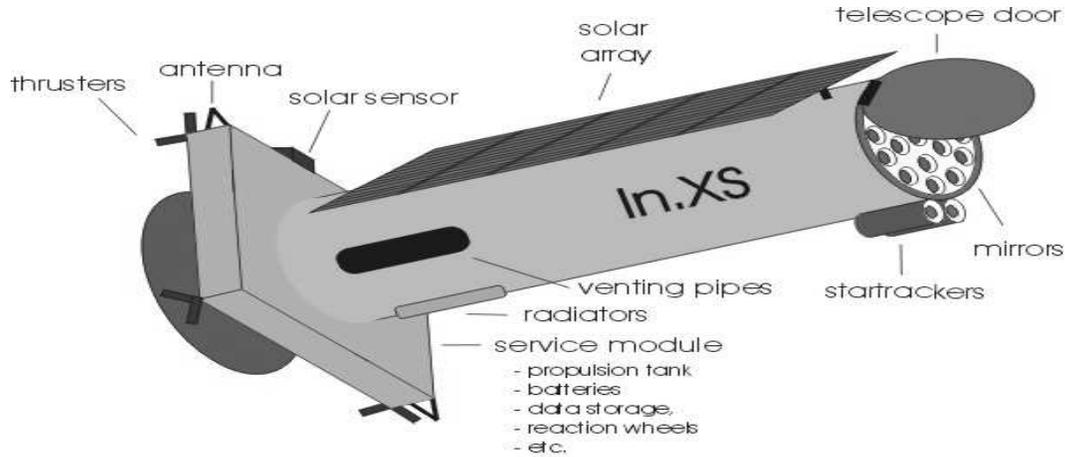,height=6cm,width=14cm}
\caption{\label{spcr} Sketch of the In.XS spacecraft.}
\end{center}
\end{figure}

A Ariane IV class (like U.S. Delta II or Japanese H-1 or Russian Cyclone) launcher should
be sufficient considering the
relatively low overall mass of the mission. However an extra long fairing configuration
of the launcher is needed to host the spacecraft, which is approximately $10 m$ long (Fig.\,\ref{spcr})
due to the required focal length.

We estimate the total energy consumption for the service module (data handling,
antennae, etc.) and payload (detectors, etc.) to be $\approx 1.1 kW$. This can
be provided by solar arrays with a total effective area
of $A_{eff} \simeq 15 m^2$ assuming $17\%$ efficiency Si cells
and $0.83 kW \cdot h$ NiCd batteries (able to endure about 5500 charge/discharge cycles
per year) for the eclipsing times\footnote{This estimate includes 20\% overhead.}.

The science objectives (Tab.\,\ref{prereq}) require the Absolute
Pointing Error (APE) to be less than $1 arcmin$, and the Relative Pointing
Error (RPE) should be kept small compared to the PSF, i.e. of the order of $10 arcsec$,
which is easily achieved with current reaction wheel techniques.

\subsection{In.XS Groundsegment}

Table\,\ref{profile} summarizes the proposed Mission Profile for the
In.XS project. It comprises three parts: the first part is dedicated
to the commencement of operation of the satellite and the instruments,
the two other parts are dealing with the observations as described in
the Mission Concept above.

\begin{table}[ht]
\caption{\label{profile} Mission Profile}
\begin{tabular}{|l|l|}
\hline
$\bullet$ \textbf{L}aunch and \textbf{E}arly \textbf{O}rbit \textbf{P}hase&\\
$\bullet$ Commissioning Phase (incl. Outgassing)&1 month\\
$\bullet$ Calibration and Performance Validation Phases&2 months\\
\hline
$\bullet$ \textbf{Whole sky survey}&1.3 years\\
\hspace*{1cm}processed data available 3 months after observation&\\
\hspace*{1cm}$\rightarrow$ Catalogue of sources (all data public domain)&\\
\hline
$\bullet$ \textbf{Pointed} follow-up \textbf{observations} of selected objects&2 years\\
\hspace*{1cm}-- Observing time divided into guaranteed time (PIs) and open proposals&\\
\hspace*{1cm}-- Announcement of Opportunity (AO) and Observation Time Allocation Commitee (OTAC)&\\
\hspace*{1cm}-- Scheduling and Observation execution&\\
\hspace*{1cm}-- Priority rights and Data distribution (PIs or public archives)&\\
\hline
\end{tabular}
\end{table}                

The Groundsegment of In.XS has the following duties: providing all facilities and
services needed for mission operation on ground, providing the data link from spacecraft 
to the ground, preparing a timeline for the operation, carrying out the monitoring, controlling
and commanding of the spacecraft and instruments, executing transmission
and the processing and archiving of the data.

The Groundsegment will consist of the following strongly interacting parts:
\begin{itemize}
\item \bf Groundstations: \rm the groundstations will be responsible for keeping contact 
with the satellite and exchange telecommands (TC, controlling) and telemetry (TM, science 
data and housekeeping), respectively. As we propose a near-equatorial
orbit for In.XS, the ESA groundstation in Kourou (French Guyana) is suitable for communication with the
satellite, a possible backup station would be the ESA facility in Malindi (Kenya).
\item \bf Mission Operations Center (MOC): \rm The MOC will be responsible for all
mission-operation related activities (safety checks for spacecraft and instruments, storing
TM, routing TM to the SOC, TC commanding). In.XS could make use of the existing facilities of 
ESA/ESOC in Darmstadt (Germany) as MOC. 
\item \bf Science Operations Center (SOC): \rm The SOC will be responsible for all science 
related activities, i.e. instrument calibration, performance monitoring and survey 
analysis. Besides, it should perform the mission planning, provide the possibility of a 
quick-look-analysis on the incoming scientific data, archive the data and make it 
available for the scientific community. The exact number and 
location(s) of required SOC(s) is left to be decided on and supported by the PI-teams.
In return, PIs will publish the all-sky catalogue and get guaranteed time during
the pointed observations phase. For data distribution, use of the ESA OPSNET should be considered.
\end{itemize}

The maximum data flow for scientific and housekeeping data was
calculated to be about $3.2 kbits \cdot s^{-1}$, which
results in $2.3 Mo$\footnote{$Mo = MegaOctet = MegaByte = 8 MegaBits = 8 \times 1024 kbits$}
per orbit ($35 Mo$ daily). To keep the costs low, we decided to use only one
telemetry downlink per day. This means, that we have to provide for 24
hours worth of onboard data-storage. To be on the safe side, total onboard
memory storage should amount to $64 Mo$ (almost 2 days, which is very reasonable when compared
to the flash-eproms Mo price rate). If one assumes that the
satellite will be visible during a $10 min$ window per orbit from the main groundstation and $50\%$ of this
time has to be used for telemetry, a minimum transmission rate of $2.4 Mbits \cdot s^{-1}s$ ($90 Mo$ per window)
would be enough by far for the daily downlink. For telecommunication with the satellite a parabolic antenna ($D = 15 m$)
at the groundstation and 2 low-gain antennae ($P = 30 W$, $f_{UL} = 7.2 GHz$, $f_{DL} = 8.5 GHz$)
onboard the satellite are required. Calculations
show that with this equipment one can reach
transmission rates well sufficient to get all the data safely
down. The calculated maximum rate for the telecommands (TC) uplink is
$64 kbits \cdot s^{-1}$ (like an InMarSat M-4 terminal),
providing sufficient margin considering the expected TC
rate of $\sim 2 kbits \cdot s^{-1}$.
One year of all-sky survey raw data archive should hence require $12.4 Go$ mass storage,
i.e. roughly one double-sided DVD.

\section{CONCLUSION AND FUTURE DIRECTIONS}

Assuming that instruments and SOC are provided by the scientific community, this project suffers
from no severe cost drivers, since it only uses standard spacecraft technology, limited groundsegment operations
and launcher configuration and requires only limited propellant capacity, thermal and attitude control.

Given its high sensitivity, $1 arcmin$ spatial resolution, $1 keV$ spectral resolution, $1 year$
all-sky survey capability in the $2 - 80 keV$ range, In.XS hence fits a typical
high technology - short term profile, suitable for an ESA flexi-mission
or a NASA ``cheaper - faster - better'' concept, which could give important results toward X-ray background
studies but also various galactic sources, AGN, clusters of galaxies, etc.

And remember: it is spelled I - n . X - S, but is pronounced ``an access'' (to the hidden universe) \ldots

\acknowledgments

The In.XS satellite is a project originating from the Alpbach 2000 summerschool,
co-organized by the Austrian and European Space Agencies (ASA, ESA). Our group
of students wish to thank again the whole Alpach organizing commitee, in particular
Johannes ``Dear Friend'' Ortner, and all our scientific tutors, R. Much, A. Parmar, M. Turner,
and also Bruno Gardini and Kevin Bennet for their precious space missions experience,
as well as all the staff of the Hotel B\"oglerhof.
   P. Marty ackowledges the financial supports of both the Centre National d'Etudes Spatiales (CNES)
and the Ecole Doctorale d'Astronomie-Astrophysique d'Ile de France (Paris University) which
allowed the attendance of this school.
   J. Schultz acknowledges the financial support of the Space Research
Programme at University of Helsinki, funded by the Academy of Finland.

Alpbach 2000 summerschool: http://www.asaspace.at/events/alpbach2000rep.html

In.XS project: http://www-station.ias.u-psud.fr/inxs


\end{document}